\documentclass[english,longauth]{aa}
\usepackage[T1]{fontenc}
\usepackage[latin1]{inputenc}
\usepackage{textcomp}
\usepackage{graphicx}
\IfFileExists{url.sty}{\usepackage{url}}
                      {\newcommand{\url}{\texttt}}
\usepackage[authoryear]{natbib}

\makeatletter

\usepackage{color}
\usepackage{hyperref}

\usepackage{units}
\usepackage{natbib}

\bibpunct{(}{)}{;}{a}{}{,}

\usepackage{babel}
\makeatother

\begin{document}
\author{A.~Nigl\inst{1}
\and W.D.~Apel\inst{2}
\and J.C.~Arteaga\inst{3}
\and T.~Asch\inst{4}
\and J.~Auffenberg\inst{5}
\and F.~Badea\inst{2}
\and L.~B\"ahren\inst{6}
\and K.~Bekk\inst{2}
\and M.~Bertaina\inst{7}
\and P.L.~Biermann\inst{8}
\and J.~Bl\"umer\inst{2,3}
\and H.~Bozdog\inst{2}
\and I.M.~Brancus\inst{9}
\and M.~Br\"uggemann\inst{10}
\and P.~Buchholz\inst{10}
\and S.~Buitink\inst{1}
\and H.~Butcher\inst{6}
\and E.~Cantoni\inst{7}
\and A.~Chiavassa\inst{7}
\and F.~Cossavella\inst{3}
\and K.~Daumiller\inst{2}
\and V.~de Souza\inst{3}
\and F.~Di Pierro\inst{7}
\and P.~Doll\inst{2}
\and R.~Engel\inst{2}
\and H.~Falcke\inst{1,6}
\and H.~Gemmeke\inst{4}
\and P.L.~Ghia\inst{11}
\and R.~Glasstetter\inst{5}
\and C.~Grupen\inst{10}
\and A.~Haungs\inst{2}
\and D.~Heck\inst{2}
\and J.R.~H\"orandel\inst{1}
\and A.~Horneffer\inst{1}
\and T.~Huege\inst{2}
\and P.G.~Isar\inst{2}
\and K.-H.~Kampert\inst{5}
\and D.~Kickelbick\inst{10}
\and Y.~Kolotaev\inst{10}
\and O.~Kr\"omer\inst{4}
\and J.~Kuijpers\inst{1}
\and S.~Lafebre\inst{1}
\and P.~\L{}uczak\inst{12}
\and M.~Manewald\inst{4}
\and H.J.~Mathes\inst{2}
\and H.J.~Mayer\inst{2}
\and C.~Meurer\inst{2}
\and B.~Mitrica\inst{9}
\and C.~Morello\inst{11}
\and G.~Navarra\inst{7}
\and S.~Nehls\inst{2}
\and J.~Oehlschl\"ager\inst{2}
\and S.~Ostapchenko\inst{2}
\and S.~Over\inst{10}
\and M.~Petcu\inst{9}
\and T.~Pierog\inst{2}
\and J.~Rautenberg\inst{5}
\and H.~Rebel\inst{2}
\and M.~Roth\inst{2}
\and A.~Saftoiu\inst{9}
\and H.~Schieler\inst{2}
\and A.~Schmidt\inst{4}
\and F.~Schr\"oder\inst{2}
\and O.~Sima\inst{13}
\and K.~Singh\inst{1}
\and M.~St\"umpert\inst{3}
\and G.~Toma\inst{9}
\and G.C.~Trinchero\inst{11}
\and H.~Ulrich\inst{2}
\and J.~van~Buren\inst{2}
\and W.~Walkowiak\inst{10}
\and A.~Weindl\inst{2}
\and J.~Wochele\inst{2}
\and J.~Zabierowski\inst{12}
\and J.A.~Zensus\inst{8}
}

\institute{ 
Department of Astrophysics, IMAPP, Radboud University Nijmegen, P.O. Box 9010, 6500 GL Nijmegen, The~Netherlands\\
\email{anigl@astro.ru.nl}\\
  \and 
Institut\ f\"ur Kernphysik, Forschungszentrum Karlsruhe, 76021~Karlsruhe, Germany\\
  \and 
Institut f\"ur Experimentelle Kernphysik, Universit\"at Karlsruhe (TH), 76021~Karlsruhe, Germany\\
  \and 
Institut f\"ur Prozessverarb. und Elektr., Forschungszentrum Karlsruhe, 76021~Karlsruhe, Germany\\
  \and 
Fachbereich Physik, Universit\"at Wuppertal, 42097~Wuppertal, Germany \\
  \and 
ASTRON, 7990 AA Dwingeloo, The Netherlands \\
  \and 
Dipartimento di Fisica Generale dell'Universit{\`a}, 10125~Torino, Italy\\
  \and 
Max-Planck-Institut f\"ur Radioastronomie, 53010~Bonn, Germany \\
  \and 
 National Institute of Physics and Nuclear Engineering, 7690~Bucharest, Romania\\
  \and 
Fachbereich Physik, Universit\"at Siegen, 57068~Siegen, Germany \\
  \and 
Istituto di Fisica dello Spazio Interplanetario, INAF, 10133~Torino, Italy \\
  \and 
Soltan Institute for Nuclear Studies, 90950~Lodz, Poland\\
  \and 
Physics Department, Bucharest University, Bucharest-Magurele, P.O. Box MG-11, RO-077125, Romania
}

\title{Direction identification in radio images of cosmic-ray air showers detected with LOPES and KASCADE}

\abstract
{}
{We want to understand the emission mechanism of radio emission from air showers to determine the origin of high-energy cosmic rays. Therefore, we study the geometry of the air shower radio emission measured with LOPES and search for systematic effects between the direction determined on the radio signal and the direction provided by the particle detector array KASCADE.}
{We produce 4D radio images on time-scales of nanoseconds using digital beam-forming. Each pixel of the image is calculated for three spatial dimensions and as a function of time. The third spatial dimension is obtained by calculating the beam focus for a range of curvature radii fitted to the signal wave front. We search this multi-dimensional parameter space for the direction of maximum coherence of the air shower radio signal and compare it to the direction provided by KASCADE.}
{The maximum radio emission of air showers is obtained for curvature radii being larger than 3 km. We find that the direction of the emission maximum can change when optimizing the curvature radius. This dependence dominates the statistical uncertainty for the direction determination with LOPES. Furthermore, we find a tentative increase of the curvature radius to lower elevations, where the air showers pass through a larger atmospheric depth. The distribution of the offsets between the directions of both experiments is found to decrease linearly with increasing signal-to-noise ratio. Significantly increased offsets and enhanced signal strengths are found in events which were modified by strong electric fields in thunderstorm clouds.}
{We conclude that the angular resolution of LOPES is sufficient to determine the direction which maximizes the observed electric field amplitude. However, the statistical uncertainty of the directions is not determined by the resolution of LOPES, but by the uncertainty of the curvature radius. We do not find any systematic deviation between the directions determined from the radio signal and from the detected particles. This result places a strong supportive argument for the use of the radio technique to study the origin of high-energy cosmic rays.}


\authorrunning{Nigl et al.}
\titlerunning{CR direction analysis with LOPES/KASCADE}
\offprints{A. Nigl}
\mail{anigl@astro.ru.nl}
\date{Received 7 December 2007 / Accepted 12 June 2008}

\maketitle

\section{Introduction}

Cosmic rays are particles penetrating the Earth's atmosphere with
energies from GeV up to hundreds of EeVs. These particles (mostly
protons) have been measured directly up to energies of PeV. Most cosmic
rays up to PeV energies are thought to be accelerated in our Galaxy
by supernova remnants. However, the origin of the particles beyond
those energies is still unclear.

Therefore, the study of high-energy cosmic rays is important to find
their sources and to understand how they were accelerated. However,
the flux drops with a steep spectrum from about one particle per year
per square meter at PeV energies, to about one particle per century
per square kilometer at energies of a hundred EeV. Because of this
low flux and a certain interaction in the Earth's atmosphere, cosmic
rays with energies beyond about a hundred PeV are measured indirectly
by a cascade of secondary particles. Billions of secondary particles
are produced in these cascades called extensive air showers and hundreds
of thousands of them reach particle detectors on the ground.

The relative arrival times of these particles allow to determine the
direction from which the primary particle entered the Earth's atmosphere.
However, the identification of the source is hampered by the deflection
of the primary particle by the magnetic field distribution in our
Galaxy. The higher the energy, the smaller the deflection of the particle;
and for very high energies larger than $10^{19}$ eV (10 EeV), the
proton cyclotron radius becomes larger than the dimension of our Galaxy
so that its trajectory possibly points back to its source.

Measurement of these cosmic rays over long periods can reveal a clustering
of events, which would enable the identification of their sources
and the determination of the particle acceleration process (\citealt{pao07}).

Cosmic-ray air showers and their origin have been studied with particle
detectors since \citet{auger39}. The position of the shower axis
on the ground is determined from the lateral distribution of particles
(\citealt{mayer92}), which decreases with distance from the shower
core; and the direction is determined on the particle arrival times
(\citealt{mayer93}).

\citet{falcke05} and \citet{codalema05} showed that radio emission
from cosmic-ray air showers can also be used to determine the energy
of the primary particle. This result revived the detection of air
showers in the radio regime, which were first observed in 1964 by
\citet{jelley65} and \citet{allan66}. The radio emission in particle
cascades is produced by electrons and positrons being deflected in
the Earth's magnetic field. Both emit beamed coherent radiation called
geosynchrotron emission (\citealt{falcke03}). Since the electromagnetic
emission is not attenuated by collisions, it reaches the ground and
therefore the longitudinal shower development can be studied by observing
the emission at several distances from the shower axis. The peak voltage
of the radio emission detectable on the ground depends approximately
linearly on the energy of the primary particle (\citealt{huege05app}).
In addition, the radio emission increases with increasing geomagnetic
angle between the cosmic-ray trajectory and the Earth's magnetic field
(\citealt{falcke05}).

A new generation of radio telescopes like LOFAR and LOPES can digitize
the signals from cosmic-ray air showers detected by many simple dipole
antennae providing them with high time, frequency, and spatial resolution.
These powerful interferometers can be used to determine the direction
of origin of the beamed and coherent emission produced in air showers.

In this article, we present the determination of the direction of
several hundreds of cosmic-ray air showers measured in the radio regime
with the LOFAR Prototype Station (LOPES, \citealt{horneffer04}) at
energies above $10^{16}$ eV. We determine the exact direction of
maximum emission by searching a small 4D radio map in the direction
provided by the triggering KArlsruhe Shower Core and Array particle
DEtector (KASCADE, \citealt{antoni03}). The resulting distance between
the radio-direction and the particle-direction is analyzed for systematic
effects.

The data acquisition and event selection is described in Sect. \ref{sec:crdirid_data}.
the direction determination is explained in Sect. \ref{sec:crdirid_max}.
The dependencies of the shower emission maximum are elaborated in
Sect. \ref{subsec:crdirid_dependences}, and the results are discussed
in Sect. \ref{sec:crdirid_sel}.

\section{Data acquisition\label{sec:crdirid_data}}

LOPES consists of 30 inverted-V-shaped dipoles placed between the
particle detector stations of the KASCADE experiment. LOPES receives
a coincidence trigger from KASCADE and stores $0.4~\unit{ms}$ of
radio data before and after its reception. The signal of each antenna
is amplified, filtered in the band from 40~MHz to 80~MHz, digitized
with 80 Megasamples per second, and written to disk. The digitally
stored time-series for each antenna is processed off-line, where we
perform absolute calibration, we reduce radio frequency interference
(RFI) by down-weighting of narrow-band lines in frequency, and form
a beam by adding up or correlating the time-shifted antennae for the
direction given by KASCADE. In addition to the direction information,
KASCADE provides parameters such as the shower core position, electron
number and muon number allowing the estimation of the primary energy.
More details on the LOPES data acquisition and data processing can
be found in \citet{nigl07crspec}.

The shower direction is reconstructed by KASCADE with an accuracy
of about 0.1 degrees at the zenith and the shower core position with
an accuracy of about 1 meter (\citealt{antoni03}), both for an event
with an electron number of $10^{6.5}$ ($\sim10^{16.5}~\unit{eV}$).
The directional accuracy improves with increasing shower size and
reduces with inclination angle. Beyond 42 degrees zenith angle, the
directional uncertainty is strongly increased due to the reduced acceptance
of the particle detectors. The mentioned accuracy of KASCADE is valid
for showers hitting the particle detector array within a centered
circle with a radius of 91 meters, so that the core is at least 10
meters inside the array. 

For this work, 664 LOPES events were pre-selected with a lower limit
on the muon number of $10^{5.2}$ ($>10^{16}~\unit{eV}$) and a cut
on the distance of the shower core from the array centre of 91 meters.
These events have been acquired from 2005-11-16 to 2006-07-23.

\section{Direction Determination\label{sec:crdirid_max}}

The phasing of an array by digital beam-forming is usually done in
direction of a source at infinite distance. However, the air shower
radio emission is produced in the Earth's troposphere by an extended
emission-region and therefore, the wavefront is not flat. As a first
approximation, a spherical wavefront was assumed for the radio emission
reaching the antennae of LOPES. The spherical wavefront is defined
by a specific curvature radius (beam focus) and a shower core position
on the ground. The resulting distance of the beam focus does not necessarily
coincide with the distance of the maximum emission in the air shower.
This needs to be investigated in shower simulations and dedicated
measurements in order to study the real shape of the shower front
and its relation to the position of the maximum radio emission and
the position of the maximum number of particles, respectively. The
relation between the maximum in the radio shower and in the particle
shower might enable the determination of the particle species from
the shower radio emission only. This topic is not part of this work,
however Monte Carlo simulations by \citet[Fig. 5]{huege05app} show
an indication of a more complex shower wavefront than a spherical
one.

\subsection{Four-dimensional radio image\label{subsec:crdirid_5dimg}}

For each cosmic-ray event, all pixels of a four-dimensional radio
image were calculated spanning azimuth, elevation, curvature radius,
and time. The image was calculated with an extension of $50\times50$
pixels, with an angular resolution of $0.2\unit{^{\circ}}$. The map
was centered on the direction given by KASCADE.

The shower core position for the spherical beam-forming was taken
from KASCADE. The curvature radius for the beam-forming algorithm
was chosen to range from $1~\unit{km}$ to $10~\unit{km}$ in 37 steps
of $250~\unit{m}$. The time resolution was the LOPES sample-time
bin of $12.5~\unit{ns}$. The time window was chosen to be $1.6~\unit{\mu s}$
(128 samples), so much larger than the length of a typical cosmic-ray
radio pulse of a few samples in time, to avoid distortion of the pulse
in the image due to edge-effects in the beam-forming process. The
signal has been integrated over the LOPES band from $43~\unit{MHz}$
to $74~\unit{MHz}$. The shower core position, which is used for the
beam-forming process remains constant at the value provided by KASCADE.
The intensity of the cosmic-ray event in the image corresponds to
the power measured in the LOPES beam. The imager performs beam-forming
by adding up the antennae and not by correlating them, as it is used
for calculating the E-field amplitude at the position of the maximum
from the image.

Oscillations in the signal shorter than the inverse of the bandwidth
are not real but they are caused by over-sampling a bandpass broadened
signal. Therefore, the final 4D image was Hanning-smoothed in time
by a weighted average of every pixel with its two neighbouring pixels.

\subsection{Emission maximum}

The emission maximum in the 4D image was located, firstly, by determining
the time of maximum emission for each slice in curvature radius by
selecting the 2D-azimuth/elevation-slice containing the pixel with
the maximum intensity. For this search in time, 41 of the 128 available
samples around the window center were scanned. Secondly, each of the
2D curvature-slices at the temporal maximum is searched for the position
of the brightest point source (FINDPOINTSOURCES function of CASA/AIPS++,
\url{www.aips2.nrao.edu}, v1.9, build 1360, \citealt{aips++96}).
A plot of an example event can be found in Fig \ref{fig:crdirid_skymaps}.
In case multiple sources were found, the direction of the strongest
one was saved and the other sources were classified as side-lobes.
As a result, the position in time and the direction in azimuth-elevation
were calculated as a function of curvature radius. 

The two-dimensional spatial shape of the cosmic-ray radio emission
in the image is mainly determined by the beam-shape (point spread
function, PSF) and depends on the layout of the selected antennae.
The PSF of the emission is assumed to be spherical at the zenith.
At lower elevations, the resolution in elevation of LOPES is reduced
by the sine of the elevation angle. A theoretical uncertainty for
the determination of the direction is obtained following \citet{taylor99}:\begin{eqnarray}
\Delta\alpha_{min} & = & ±\frac{1}{2}\frac{\epsilon_{N}}{\epsilon_{P}}\frac{\lambda}{D\,\mbox{cos}\theta}\label{eq:crdirid_angres}\end{eqnarray}

Here $\Delta\alpha_{min}$ is the minimum statistical uncertainty
of the determined direction, $\epsilon_{P}$ is the electric field
strength of the detected pulse, $\epsilon_{N}$ is the noise in the
LOPES beam, $\lambda$ is the wavelength, $D$ is the diameter of
the LOPES antenna array, and $\theta$ is the zenith angle. The angular
resolution $\vartheta=\lambda/D\,\mbox{cos}\theta$ or beam width
of LOPES is $2.2\unit{^{\circ}}$ for the zenith ($\lambda=5~\unit{m}$,
$D\,\mbox{cos}\theta=130~\unit{m}$). An example-event of LOPES at
an estimated primary energy of 50 PeV, an electron number of $10^{6.7}$,
and a zenith angle of $13.0\unit{^{\circ}}$ (for LOPES and KASCADE)
was obtained with a E-field peak signal-to-noise ratio (SNR) of $\epsilon_{P}/\epsilon_{N}=8.5$
resulting in an uncertainty of $0.13\unit{^{\circ}}$, which is about
the same accuracy as mentioned for KASCADE.

\subsection{Dependencies of the measured electric field\label{subsec:crdirid_dependences}}

\begin{figure}
\begin{centering}
\includegraphics[width=1\columnwidth]{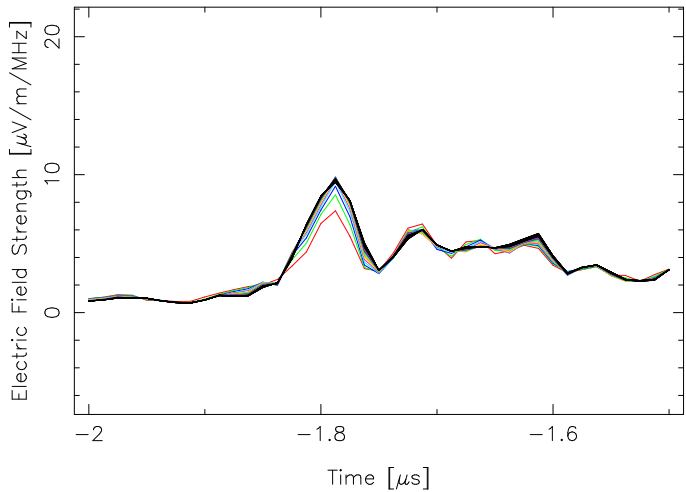} 
\par\end{centering}

\caption{\label{fig:crdirid_dep_time}Maximum electric field amplitude of the
image as a function of time for different curvature radii. The amplitude
of the pulse at $-1.8~\unit{\mu s}$ increases with beam-forming curvature
radii (as shown for another event in Fig. \ref{fig:crdirid_dep_dist}).}

\end{figure}

In the time domain, the air shower pulse is expected at approximately
$1.8~\unit{\mu s}$ before the trigger from KASCADE arrives at the
antennae (\citealt{horneffer04}). Fig. \ref{fig:crdirid_dep_time}
shows the intensity maximum of the image-slices as a function of time
(x-axis) and curvature radius (different curves). The curves for the
different curvature radii from 1~km to 10~km show that the intensity
in the signal at $-1.8~\unit{\mu s}$ is increasing until the best
fit is found, whereas the intensity of the background noise stays
the same.

\begin{figure}
\begin{centering}
\includegraphics[width=1\columnwidth]{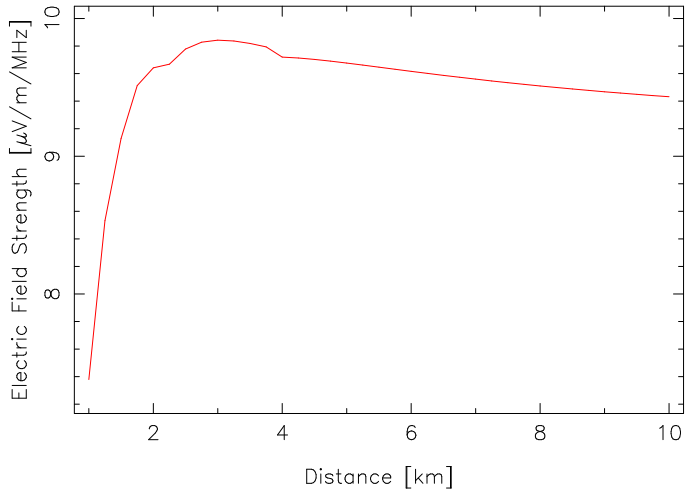} 
\par\end{centering}

\caption{\label{fig:crdirid_dep_dist}Maximum electric field amplitude of the
image as a function of beam-forming radius (distance from the observer
to the beam focus) for an event with an estimated primary energy of
$3\times10^{17}~\unit{eV}$.}

\end{figure}

The intensity of the maximum source in each skymap increases steeply,
when increasing the curvature radii from 1~km to about 3~km, which
is due to increasing coherence of the cosmic-ray air shower pulses
in the individual antenna signals. For larger curvature radii, the
intensity either keeps increasing asymptotically or a maximum is found,
as in Fig. \ref{fig:crdirid_dep_dist}. For the plotted event the
intensity peaks at 3~km.

\begin{figure}
\begin{centering}
\includegraphics[width=1\columnwidth]{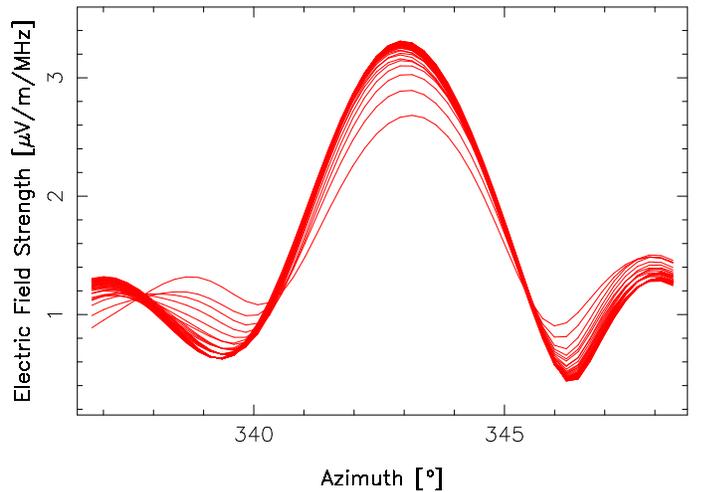}
\par\end{centering}

\vspace{20pt}

\begin{centering}
\includegraphics[clip,width=1\columnwidth]{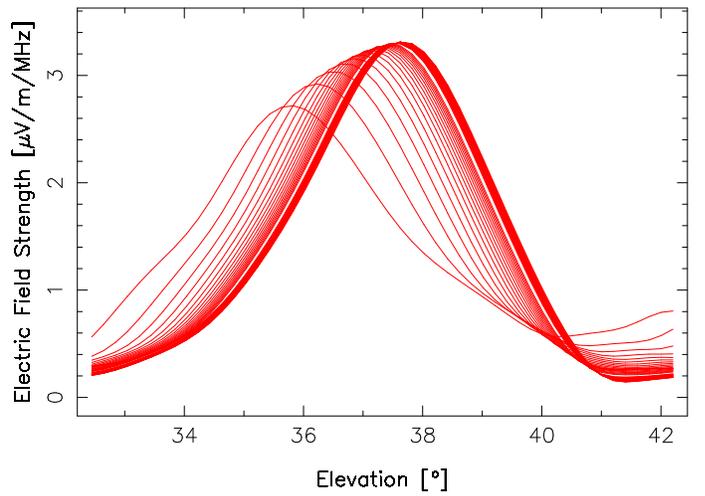}
\par\end{centering}

\caption{\label{fig:crdirid_dep_dir}Electric field amplitude as a function
of azimuth (top) and elevation (bottom) at the time of maximum emission
in the image for different curvature radii. The amplitude of the curves
increases with beam-forming radii (as shown for another event in Fig.
\ref{fig:crdirid_dep_dist}). The radius of curvature is increased
by steps of 250~m.}

\end{figure}

The intensity as a function of azimuth, elevation, and curvature radius
(different curves) is shown in Fig. \ref{fig:crdirid_dep_dir}. The
different curves show with increasing curvature radius the increase
in the intensity of the pulse, while the intensity of the side-lobes
next to the pulse stay the same or are reduced in intensity.

In addition to the change in the intensity, in the plot with intensity
versus elevation (bottom panel of Fig. \ref{fig:crdirid_dep_dir}),
the position of the pulse-maximum shifts with increasing curvature
radius to higher elevations by about two degrees. We attribute this
to the offset between the phase center of the LOPES array and the
shower core position as provided by KASCADE. The larger the distance
between both, the more the position of the emission maximum changes,
when varying the curvature radius for the beam-forming.

\begin{figure}
\begin{centering}
\includegraphics[clip,width=1\columnwidth]{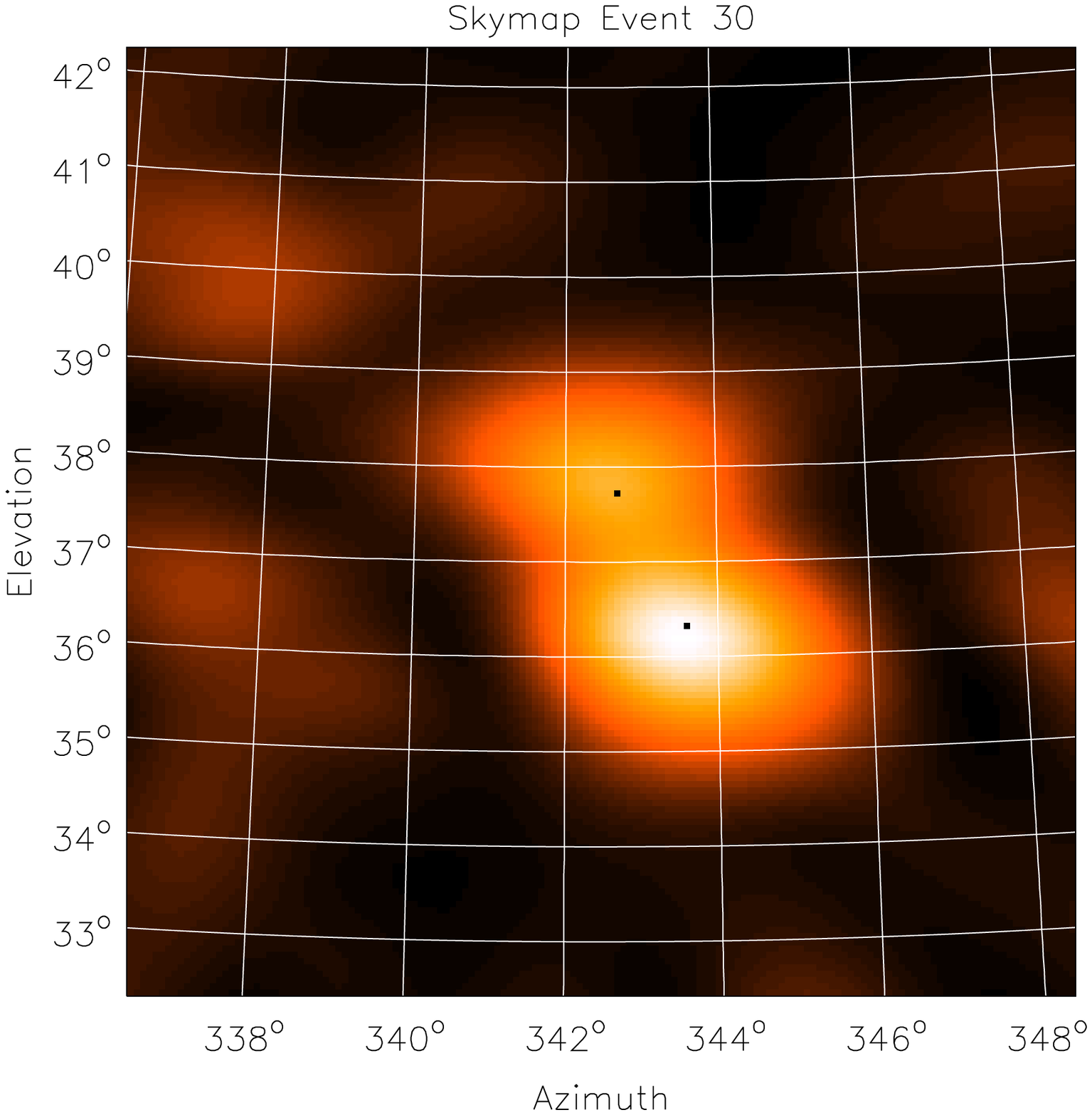}
\par\end{centering}

\begin{centering}
\includegraphics[width=1\columnwidth]{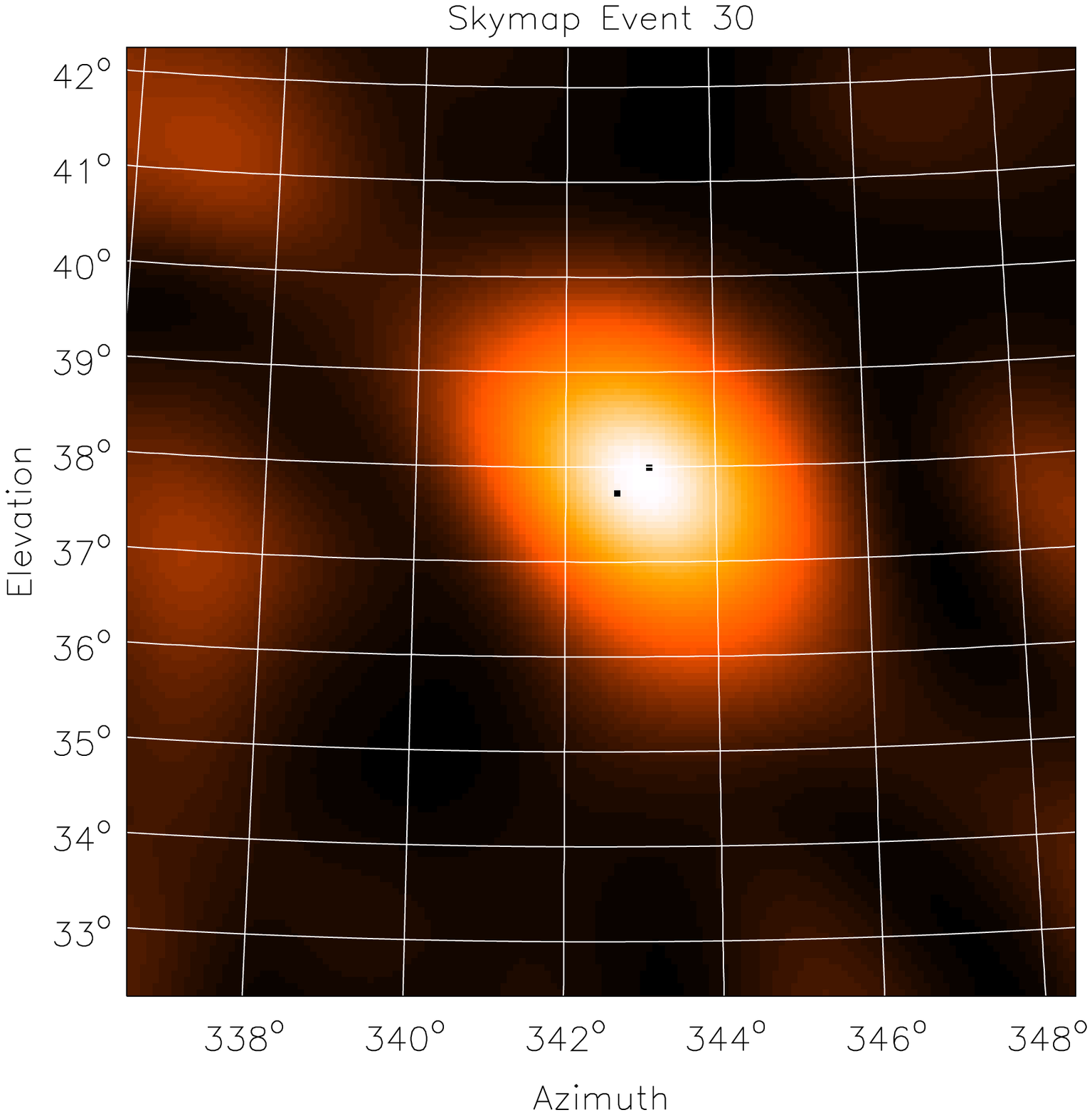}
\par\end{centering}

\caption{\label{fig:crdirid_skymaps}Example of radio skymaps for one LOPES
event with original sampling (top) and with a half-sample shift applied
to all antennae signals before beam-forming (bottom). The split maximum
in the top plot is not real and merges to one maximum with higher
temporal resolution. The skymaps show the air shower radio emission
in stereographic projection (STG, \citealt{calabretta02}) with a
black dot at the maximum. The second black dot in the center of the
map and offset from the maximum indicates the direction provided by
KASCADE.}

\end{figure}

This beam-forming effect was tested by artificially introducing an
extra delay by half a sample to all antenna data. This subsample shift
is equivalent to a half-sample interpolation of the data, when correcting
the time-axis by half a sample in the opposite shifting direction.
The result of this method is displayed in Fig. \ref{fig:crdirid_skymaps},
which is showing the skymaps for the unmodified and for the shifted
data, respectively. It turns out, that the interpolation of the antenna
data, from a sample time of 12.5~ns to 6.25~ns, can further optimize
the electric field amplitude of the air shower signal found in the
beam. This is due to the sampling of the LOPES antennae signals in
the second Nyquist-zone from 40~MHz to 80~MHz with a sample frequency
of 80~MHz. Second Nyquist-sampling preserves all signal information,
which can be recovered by digital upsampling with, in our case, a
factor of two.

We effectively did this by analyzing the interpolated data. In addition
to an increase in the electric field amplitude, the pulse shifts in
position, as observed when optimizing the curvature radius. Therefore,
we conclude that the upsampling is necessary for the determination
of the correct electric field amplitude. As expected, this will increase
the accuracy of LOPES in field strength and direction estimate.

\section{Event selection\label{sec:crdirid_sel}}

The 664 pre-selected events (see Sect. \ref{sec:crdirid_data}) were
processed as described above, including the half-sample shift. Out
of these 664 events, 232 were selected having a minimum E-field peak
SNR of 6. The cut on the SNR of the events has been applied since
for low SNR the measured noise from the KASCADE particle detectors
dominates the air shower emission, which leads to a wrong fit when
searching for the pulse profile in the time window of the 4D image.

The plot in Fig. \ref{fig:crdirid_dep_time} shows this window and
the air shower pulse lies at $-1.8~\unit{\mu s}$, with increased
curvature radius rising above the particle detector noise of KASCADE
at $-1.75~\unit{\mu s}$ to $-1.6~\unit{\mu s}$. For events with
a SNR below 6 the cosmic-ray radio pulse does not exceed the emission
from KASCADE and the wrong signal is fitted. A third selection criterion
on the coherence of the signal, obtained from the cross-correlation
beam, and a fourth criterion on the elevation angle not to be smaller
than $15\unit{^{\circ}}$ reduced the number of the selected events
further to 62.

The cross-correlation beam is formed by correlating all paired combinations
of the shifted antenna signals. The third selection criterion required
the ratio of the electric field amplitudes of the cross-correlation
beam and the total power beam to be larger than 0.73, which is one
sigma above the average fraction of all events. This selection favors
inclined events, since RFI from the KASCADE detectors is reduced,
due to their reduced acceptance to larger zenith angles (see Fig.
\ref{fig:crdirid_arrows} and \citealt{petrovic06}). This way, events
which consist of RFI but have a SNR above 6 are rejected. Such events
were found to have small curvature radii, since most RFI is emitted
locally. We note, that by this criterion also events were rejected
from air showers which did not reach the required level of signal-coherence.

The rejection of RFI events can be tested by plotting a histogram
of the distribution of the curvature radii. It can be seen in Fig.
\ref{fig:crdirid_distHIST} that indeed all events which had an emission
maximum at a curvature radius below 2000~m are not part of the filtered
distribution, these events are considered as RFI events.

Among the 62 selected events, 17 events have an electric field strength
of more than twice the expected field strength estimated from the
muon number, geomagnetic angle, and shower axis distance (\citealt{horneffer07}).
The measured electric field strength of the 62 selected events is
plotted versus the directional offset in Fig. \ref{fig:crdirid_phVSdiroff}.
In this plot, the 17 enhanced events are marked with squares and the
5 encircled events were confirmed to be recorded during thunderstorm
activity with lightnings in the vicinity of LOPES. The error bars
in the plot give the minimum uncertainty calculated with Eq. \ref{eq:crdirid_angres}.

\section{Results and discussion\label{sec:crdirid_results}}

The 5 encircled events of Fig. \ref{fig:crdirid_phVSdiroff} are candidates
for enhanced shower radio emission by electric fields in thunderstorm
clouds (\citealt{buitink07}). These five thunderstorm events show
large deviations in position, which suggests that the net force of
the electric field in thunderstorm clouds can change the direction
in which the emission is beamed. One of the 5 encircled events did
not exceed the expected field strength by a factor of two, but had
a strong offset in direction. The other 13 enhanced events could not
be associated with thunderstorm activity, however, an effect by an
increased geoelectric field in a cumulonimbus cloud without lightning
is not excluded. Therefore, the enhanced events and the thunderstorm
events were not taken into account for the following interpretation
of the results. The remaining 44 events show an approximately linear
decrease in directional offset with increasing signal strength.

\subsection{Curvature radius}

As mentioned in Sect. \ref{subsec:crdirid_dependences}, the intensity
detected in the pulses increases quickly with curvature radius. In
fact, 90\% of the maximum intensity is obtained for the selected 44
events at (2600\textpm{}800)~m and the curvature radii at the maxima
are (7000\textpm{}2000)~m (one sigma statistical uncertainties).
The statistical uncertainty for the radii at the maxima is large,
because the change in the intensity at large curvature radii is small
and cannot be determined more accurately with an array of the size
of LOPES. An approximation for the statistical uncertainty of the
curvature radius can be obtained adapting Eq. \ref{eq:crdirid_angres}:\begin{eqnarray}
\left(\frac{\Delta\rho}{\rho}\right)_{min} & = & ±\frac{\rho}{D}\frac{\epsilon_{N}}{\epsilon_{P}}\frac{\lambda}{D\,\mbox{cos}\theta}.\label{eq:crdirid_radiuserror}\end{eqnarray}

Here $\Delta\rho$ is the uncertainty of the radius of curvature \textbf{$\mathbf{\mathnormal{\rho=7000~\unit{m}}}$}
and for $D=130~\unit{m}$, $\lambda=5~\unit{m}$, $\epsilon_{N}/\epsilon_{P}=1/16$,
and $\theta=40\unit{^{\circ}}$ results in \textpm{}1400~m, which
is consistent with the above measured \textpm{}2000~m.

The histogram of the curvature radii in Fig. \ref{fig:crdirid_distHIST}
shows that a few events accumulate at 10~km indicating that those
events could be fitted with even larger curvature radii. However,
on average, the intensity does not change significantly increasing
the curvature radius over the last 2~km, and therefore the wavefront
is considered to be nearly planar. A tendency to larger curvature
radii is observed for events coming from lower elevation angles, which
travelled a longer path through the Earth's atmosphere (see Fig. \ref{fig:crdirid_distVSze}).

Furthermore, we showed that the variation of the curvature radius
can change the direction of the pulse in elevation. We found that
the angular offset in elevation from the optimum position as a function
of curvature radius is on average zero for the 44 selected events
and therefore does not introduce a systematic effect. However, the
spread must be considered as a statistical uncertainty of $±0.3\unit{^{\circ}}$
for the directions determined with LOPES.

\subsection{Upsampling}

Additionally, we showed in Sect. \ref{subsec:crdirid_dependences},
that shifts in the directions occur when maximizing the intensity
of the pulse by interpolating the antenna signals. The change in the
positions introduced by the half-sample shifts amounts to $(0.7±0.9)\unit{^{\circ}}$,
for the 44 selected events. This result determines the improvement
of the directional accuracy of LOPES. Effectively, the digitization
step error was reduced, after the signal was upsampled.

The optimization of the beam-forming by the upsampling (half-sample
shift) did not significantly increase the electric field amplitude
of the selected events with $(1±3)~\unit{µV~m^{-1}~MHz^{-1}}$ compared
to the values determined with the standard analysis software (\citealt{bahren06}).
However, it reduced the number of events with more than one source
in the optimized skymap by 30\%. Thus, the intensity of the environmental
background noise in the side-lobes of the event beams was reduced
by the optimized beam-forming.

\subsection{Systematics}

\begin{figure}
\begin{centering}
\includegraphics[width=1\columnwidth]{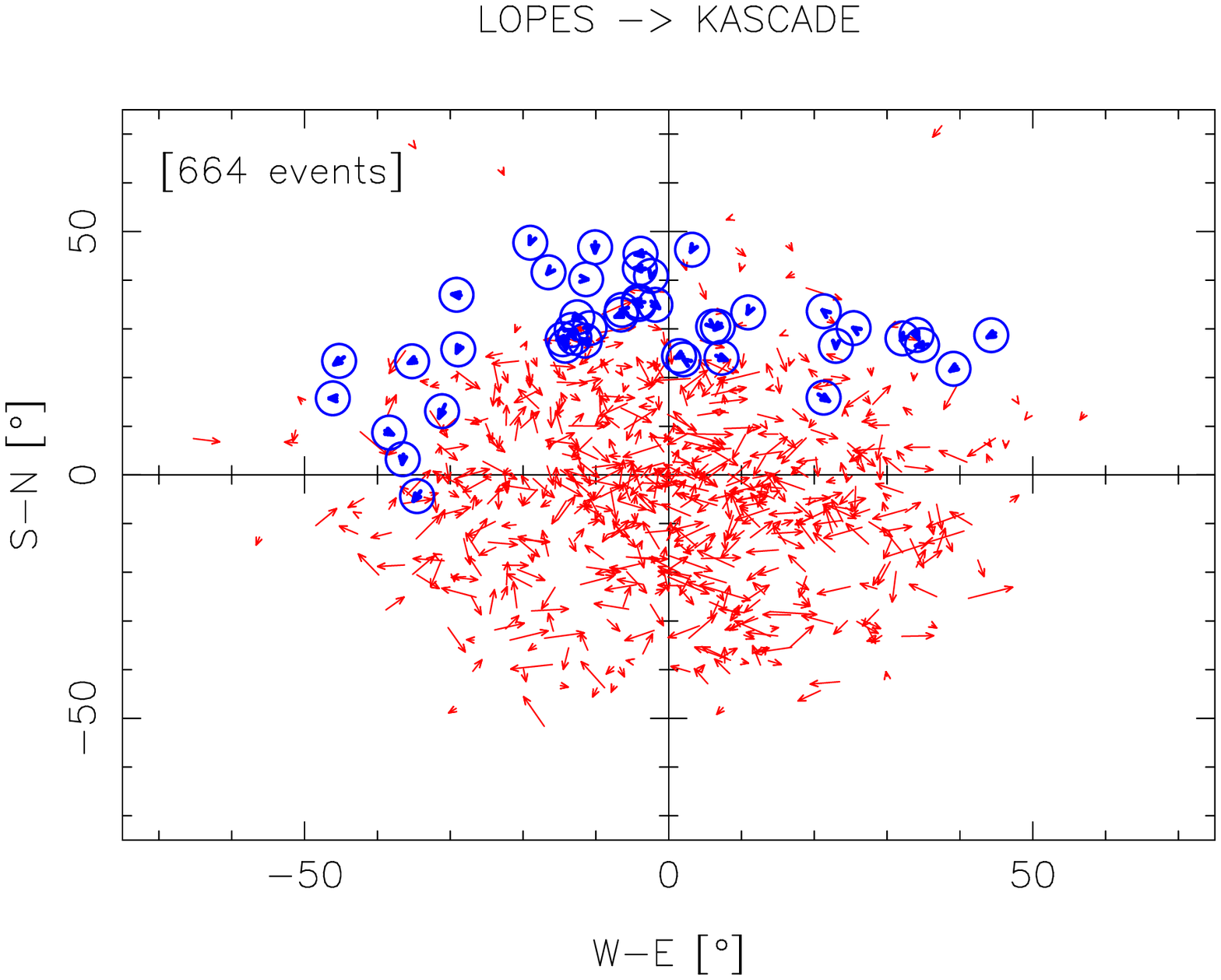} 
\par\end{centering}

\caption{\label{fig:crdirid_arrows}This plot shows an arrow for each event
connecting the position of LOPES and KASCADE projected onto the sky
with the zenith in the origin. The azimuth angle rotates clock-wise
starting from North pointing upwards. The thin arrows indicate all
664 pre-selected events and the thick encircled arrows highlight the
44 selected events.}

\end{figure}

\begin{figure}
\begin{centering}
\includegraphics[width=1\columnwidth]{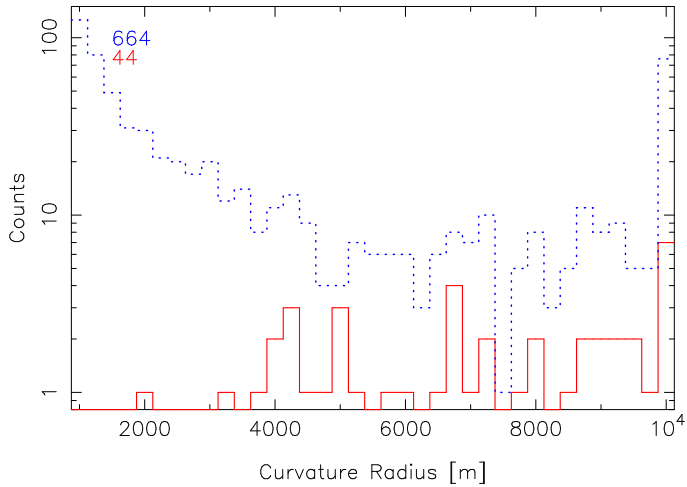} 
\par\end{centering}

\caption[Histogram of the curvature radii.]{\label{fig:crdirid_distHIST}Histogram of the number of events as
a function of curvature radius. The upper (dotted-blue) distribution
shows all 664 pre-selected events and the lower (solid-red) distribution
shows the selected 44 events. The bin size is 250 m in curvature radius.
}

\end{figure}

\begin{figure}
\begin{centering}
\includegraphics[width=1\columnwidth]{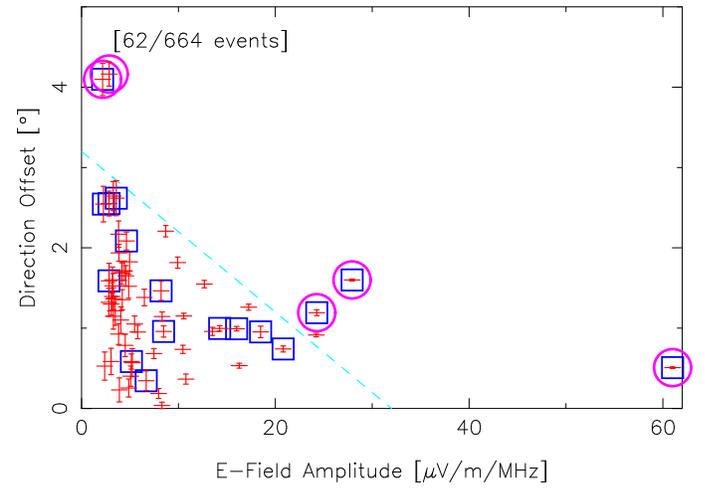} 
\par\end{centering}

\caption{\label{fig:crdirid_phVSdiroff}Absolute angular separation between
the LOPES and the KASCADE position as a function of electric field
amplitude. The encircled events are possibly enhanced in field strength
in thunderstorm clouds at the time of the observation. The events
marked with squares have a measured field strength of more than twice
the expected field strength estimated from the muon number, geomagnetic
angle, and shower axis distance (\citealt{horneffer07}). The lower
limit on the statistical uncertainty in the direction of each event
is plotted as an error bar (according to Eq. \ref{eq:crdirid_angres}).}

\end{figure}

\begin{figure}
\begin{centering}
\includegraphics[width=1\columnwidth]{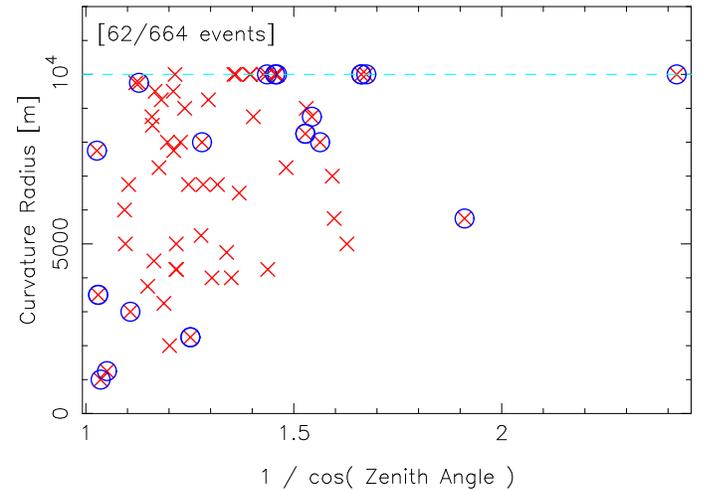} 
\par\end{centering}

\caption{\label{fig:crdirid_distVSze}Curvature radius as a function of the
inverse cosine of the zenith angle for all 62 events. The enhanced
and thunderstorm events are encircled in this plot. The dashed line
at 10~km indicates the maximum curvature radius that was fitted to
the wavefronts.}

\end{figure}

Fig. \ref{fig:crdirid_arrows} shows arrows pointing from the position
reconstructed by LOPES to the position reconstructed by KASCADE. The
arrows for all events are isotropically distributed in the circular
area that covers the sky. However, the arrows of the 44 events (selected
on their E-field peak SNR) show a clear preference for coming from
the North. Due to the inclination of the magnetic field of $65\unit{^{\circ}}$
at the latitude of LOPES of $49\unit{^{\circ}}$, the geomagnetic
angles are larger to the North than to the South and therefore a larger
signal is produced in the air shower by the geomagnetic effect.

The components of the arrows in the East-West direction and the North-South
direction are plotted in Fig. \ref{fig:crdirid_res_EW_NS}. Both data
sets follow a normal distribution with a full width at half maximum
(FWHM) of $1.2\unit{^{\circ}}$ ($\chi_{red}^{2}=2.37$) for the North-South
direction and of $0.6\unit{^{\circ}}$ ($\chi_{red}^{2}=0.88$) for
the East-West direction. These values are consistent with the absolute
offset of $(1.3±0.8)\unit{^{\circ}}$ (see next section). Furthermore,
both distributions have an average consistent with zero and therefore,
no systematic offset is found in the directions determined by LOPES
and KASCADE.

\subsection{Statistics}

The obtained offsets between the directions provided by KASCADE and
the directions determined from the LOPES radio data for the 44 selected
events result in $(1.3±0.8)\unit{^{\circ}}$. The histogram of these
offsets and of all 664 events peak at the same bin of $1.5\unit{^{\circ}}$
suggesting that this is the sum of the average uncertainty for the
direction determination of both experiments for the whole range of
elevation angles. All events having larger offsets are dominated by
increased noise from the particle detectors and they are not part
of the selected 44 events (see Fig \ref{fig:crdirid_diroffHIST}).

\begin{figure}
\begin{centering}
\includegraphics[width=1\columnwidth]{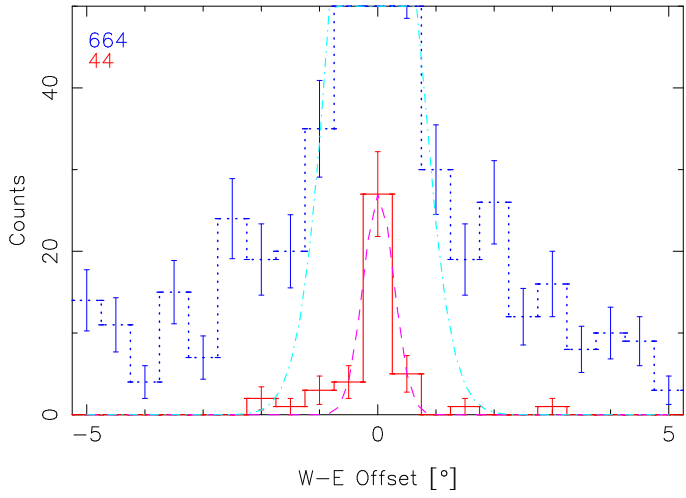}
\par\end{centering}

\vspace{20pt}

\begin{centering}
\includegraphics[width=1\columnwidth]{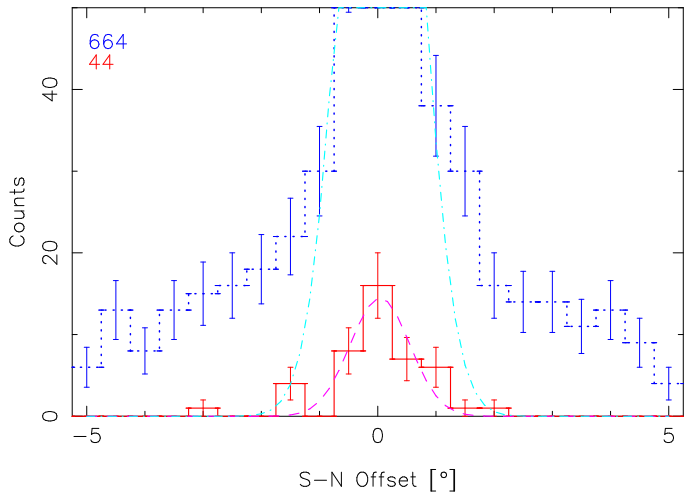}

\caption{\label{fig:crdirid_res_EW_NS}Histograms of the angular offset in
the air shower core position on the sky at the distance of the curvature
radius of the air shower emission maximum, as found in the LOPES radio
signal and as determined with the direction provided by KASCADE. The
distribution in East-West offset is plotted at the top and in North-South
direction at the bottom. Each plot shows the distribution for all
664 events (dotted) and for the 44 selected events (solid), both fitted
with a Gaussian (dash-dotted and dashed).}

\par\end{centering}
\end{figure}

\begin{figure}
\begin{centering}
\includegraphics[width=1\columnwidth]{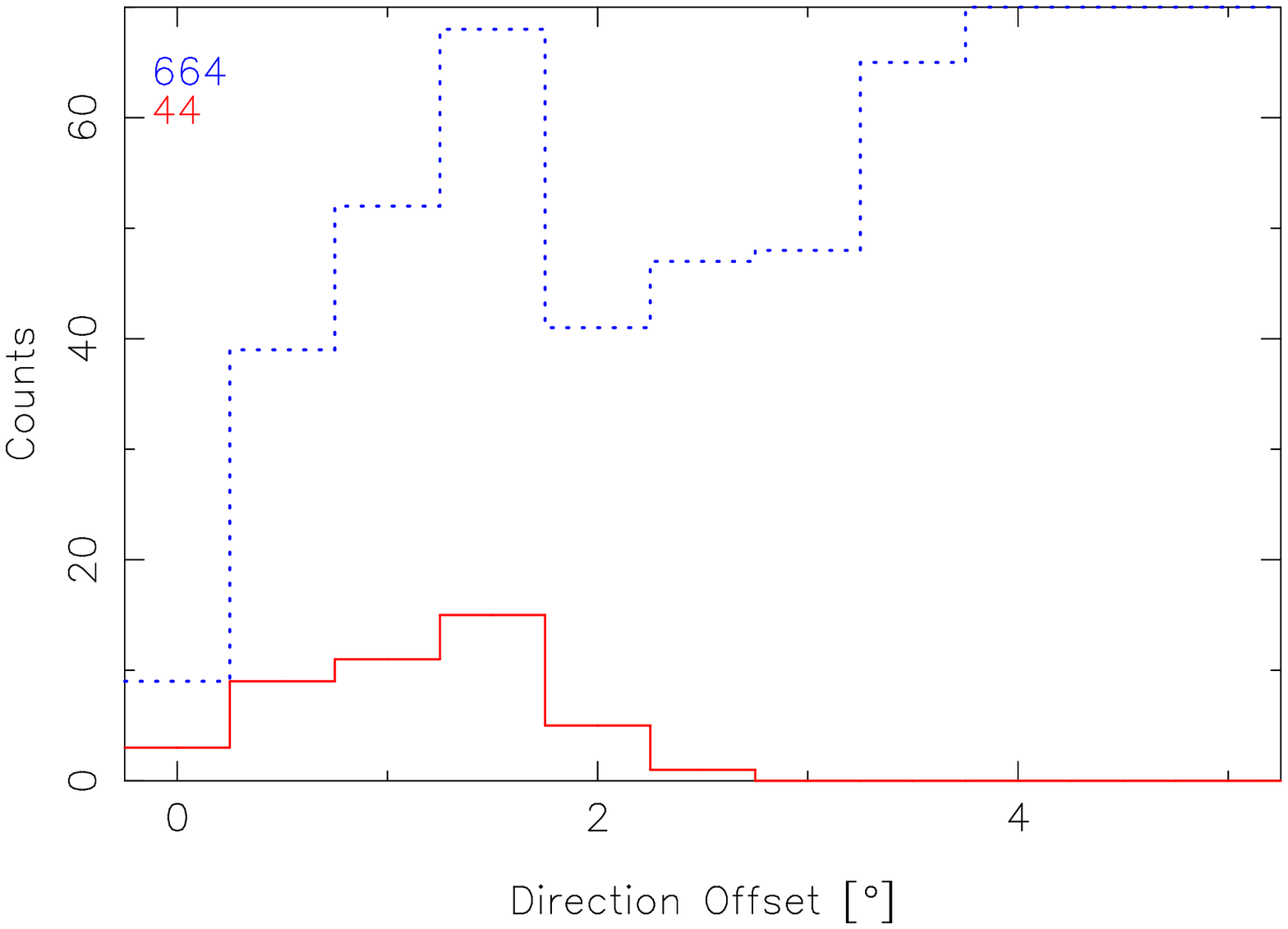} 
\par\end{centering}

\caption{\label{fig:crdirid_diroffHIST}Distribution of events as a function
of directional offset between LOPES and KASCADE. The upper (dotted)
distribution shows all 664 pre-selected events and the lower (solid)
distribution shows the selected 44 events with the maximum at $(1.3±0.8)\unit{^{\circ}}$.
The bin size is $0.5\unit{^{\circ}}$.}

\end{figure}

The clear peaks in the curves for all 664 events of Fig. \ref{fig:crdirid_res_EW_NS}
and Fig. \ref{fig:crdirid_diroffHIST} indicate that the direction
was successfully determined within $1.3\unit{^{\circ}}$ for more
than the 44 events. However, those events could also consist of RFI.
The distinction between cosmic-ray events and RFI events could be
improved by a better knowledge of the position and shape of the air
shower pulse in time. This knowledge would allow a fit of the pulse
in Fig. \ref{fig:crdirid_dep_time}, which would enable the direction
identification of events detected with less intensity than the corresponding
KASCADE noise in the LOPES beam.

The 44 selected events have an average zenith angle of $40\unit{^{\circ}}$
and the average electron number detected by KASCADE is $10^{6.8}$.
The accuracy of KASCADE for an air shower with this electron number
is $±0.2\unit{^{\circ}}$. This value is corrected for the reduced
acceptance of the particle detectors by dividing with the sine of
the zenith angle of $40\unit{^{\circ}}$. Therefore, we conclude that
the measured offsets are dominated by the accuracy of the direction
determination with LOPES of about $±1.1\unit{^{\circ}}$.

This accuracy of LOPES lies above the statistical uncertainty determined
with Eq. \ref{eq:crdirid_angres} of $±0.2\unit{^{\circ}}$, since
the fit of a curved wavefront with a curvature radius $\rho$ introduces
an additional uncertainty:\begin{eqnarray}
\Delta\alpha_{tot} & = & ±\frac{1}{2}\frac{\epsilon_{N}}{\epsilon_{P}}\frac{\lambda}{D\mbox{\, cos}\theta}+f\left(\rho\right).\label{eq:crdirid_direrror}\end{eqnarray}

\section{Conclusions}

The 4D mapping of radio emission is a powerful new tool for the study
of radio sources within the Earth's atmosphere, in particular cosmic-ray
air showers, because of the information in the third spatial dimension
provided by the distance of the beam-forming focus. We used an array
with only 30 single dipoles and succeeded to map out the radio emission
from air showers, to search for the position on the sky, and to fit
the curvature of the detected wave front of the shower pulse. For
this work, we used KASCADE as a reference. The mapping-tool can also
be used for astrophysical imaging of fast pulses, since images can
be calculated on sample time resolution. Furthermore, the third spatial
dimension of the images allows to distinguish between the location
of all kinds of transient events occurring inside and outside the
Earth's atmosphere.

We found that the position of the emission maximum in the map, calculated
in the beam-forming process, is sensitive to the array layout and
the phase errors of the antenna electronics and therefore changes
when beam-forming parameters such as the curvature radius or the filtered
band are varied. We optimized the beam-forming by upsampling of the
antenna signals, which strongly improved the accuracy of the direction
determination.

We conclude that the angular resolution of LOPES is sufficient to
localize the air shower axis and to maximize the received electric
field amplitude. However, the accuracy for the direction determination
is determined by the uncertainty of the curvature radius. An exact
determination of the shape of the shower wavefront seems to be necessary
to further improve the localization of the shower radio maximum in
the beam-forming process. Furthermore, the directional accuracy could
possibly be improved by larger baselines and by calculating every
pixel of the searched skymap with the cross-correlation beam (sum
of all antenna pair correlations) and by further suppressing the KASCADE
noise. Also, more phase-stability of the electronics and a larger
antenna layout optimized for beam-forming in all azimuthal directions
is desirable. 

We found that the best fit in the curvature radius increases for events
coming from the horizon after passing through a larger atmospheric
depth. Furthermore, we found an approximately linear decrease in the
directional offsets between the LOPES positions and the KASCADE positions
with increasing signal strength.

Furthermore, we found a few events with field strengths enhanced by
more than twice the expected field estimated from the muon number,
geomagnetic angle, and shower axis distance. Some of these events
were associated with thunderstorms in the vicinity of the telescope,
which can enhance the emission through strong electric fields in clouds.
These events showed the largest offsets between the position measured
with the radio antennae of LOPES and the particle detectors of KASCADE.
This very important result suggests that the charged particles of
air showers experience additional deflection by electric fields in
thunderstorm clouds. Thus, we measured for the first time a geoelectric
effect in addition to the geomagnetic effect.

The offsets between the directions measured with both instruments
were analyzed for dependencies on the measured electric field strength,
the curvature radius, the azimuth angle, the elevation angle and the
geomagnetic field angle. Due to the geomagnetic effect and the geoelectric
effect discussed here, one could expect a net directional offset between
the muons, the electrons and the electromagnetic emission at the shower
maximum. However, the analyzed sample of LOPES events did not show
any significant systematic effect for the mentioned dependencies.

This result adds another strong argument for the usage of the radio
detection technique for the study of the arrival directions of high-energy
cosmic rays. An interesting next step will be to investigate with
simulations how the distance of the beam focus (curvature radius)
relates to the maximum of the particle shower.

\emph{\small Acknowledgments.}{\small{} Andreas Nigl gratefully acknowledges
a grant from ASTRON which made this work possible. LOPES was supported
by the German Federal Ministry of Education and Research. The KASCADE-Grande
experiment is supported by the German Federal Ministry of Education
and Research, the MIUR and INAF of Italy, the Polish Ministry of Science
and Higher Education and the Romanian Ministry of Education and Research.}{\small \par}

\bibliographystyle{aa}
\bibliography{anigl}

\end{document}